\documentclass[12pt,preprint]{aastex}

\def\lsim{\raise0.3ex\hbox{$<$}\kern-0.75em{\lower0.65ex\hbox{$\sim$}}}
\def\gsim{\raise0.3ex\hbox{$>$}\kern-0.75em{\lower0.65ex\hbox{$\sim$}}}

\shorttitle{The rise of SN~2014J in M82}
\shortauthors{Goobar et. al.}

\begin{document}


\title{The rise of SN~2014J in the nearby galaxy M~82}


\author{%
A.~Goobar\altaffilmark{1},
J.~Johansson\altaffilmark{1},
R.~Amanullah\altaffilmark{1},
Y.~Cao\altaffilmark{2},
D.~A.~Perley\altaffilmark{2,3},
M.~M.~Kasliwal\altaffilmark{4},
R.~Ferretti\altaffilmark{1},
P.~E.~Nugent\altaffilmark{5,6},
C.~Harris\altaffilmark{5,6},
A.~Gal-Yam\altaffilmark{7},
E.~O.~Ofek\altaffilmark{7},
S.~P.~Tendulkar\altaffilmark{2},
M.~Dennefeld\altaffilmark{8},
S.~Valenti\altaffilmark{9,10},
I.~Arcavi\altaffilmark{9,11},
D.~P.~K.~Banerjee\altaffilmark{12},
V.~Venkataraman\altaffilmark{12},
V.~Joshi\altaffilmark{12},
N.~M.~Ashok\altaffilmark{12},
S.~B.~Cenko\altaffilmark{13,14},
R.~F.~Diaz\altaffilmark{15},
C.~Fremling\altaffilmark{16},
A.~Horesh\altaffilmark{7},
D.~A.~Howell\altaffilmark{9,10},
S.~R.~Kulkarni\altaffilmark{2},
S.~Papadogiannakis\altaffilmark{1},
T.~Petrushevska\altaffilmark{1},
D.~Sand\altaffilmark{17},
J.~Sollerman\altaffilmark{16},
V.~Stanishev\altaffilmark{18},
J.~S.~Bloom\altaffilmark{6},
J.~Surace\altaffilmark{19},
T.~J.~Dupuy\altaffilmark{20},
M.~C.~Liu\altaffilmark{21}
}
\altaffiltext{1}{The Oskar Klein Centre, Physics Department, Stockholm University,
    Albanova University Center, SE 106 91 Stockholm, Sweden}
\altaffiltext{2}{Cahill Center for Astrophysics, California Institute of Technology, Pasadena, CA 91125, USA}
\altaffiltext{3}{Hubble Fellow}
\altaffiltext{4}{Observatories of the Carnegie Institution for Science, 813 Santa Barbara St, Pasadena CA 91101, USA}
\altaffiltext{5}{Computational Cosmology Center, Computational Research Division, Lawrence Berkeley National Laboratory, 1 Cyclotron Road MS 50B-4206, Berkeley, CA, 94720, USA}
\altaffiltext{6}{Department of Astronomy, University of California Berkeley, B-20 Hearst Field Annex \# 3411, Berkeley, CA, 94720-3411, USA}
\altaffiltext{7}{Dept. of Particle Physics and Astrophysics, Weizmann Institute of Science, Rehovot, 76100, Israel}
\altaffiltext{8}{CNRS, Institut d'Astrophysique de Paris (IAP) 
and University P. et M. Curie (Paris 6), 98bis, Boulevard Arago F-75014 Paris, France}
\altaffiltext{9}{Las Cumbres Observatory Global Telescope Network, 6740 Corona Drive, Suite 102, Goleta, CA 93117, USA}
\altaffiltext{10}{Department of Physics, University of California, Santa Barbara, Broida Hall, Mail Code 9530, Santa Barbara, CA 93106-9530, USA}
\altaffiltext{11}{Kavli Institute for Theoretical Physics, University of California, Santa Barbara, CA 93106, USA}
\altaffiltext{12}{Physical Research Laboratory, Ahmedabad 380 009, India}
\altaffiltext{13}{Astrophysics Science Division, NASA Goddard Space Flight Center, Mail Code 661, Greenbelt, MD, 20771, USA}
\altaffiltext{14}{Joint Space Science Institute, University of Maryland, College Park, MD 20742,USA}
\altaffiltext{15}{Observatory of Geneva, University of Geneva 51 Chemin des Maillettes, 1290 Sauverny, Switzerland}
\altaffiltext{16}{The Oskar Klein Centre, Astronomy Department, Stockholm University, Albanova University Center, SE 106 91 Stockholm, Sweden}
\altaffiltext{17}{Physics Department, Texas Tech University, Lubbock, TX 79409, USA}
\altaffiltext{18}{CENTRA-Centro Multidisciplinar de Astrof\'{\i}sica, IST, Avenida Rovisco Pais, 1049-001 Lisboa, Portugal}
\altaffiltext{19}{Infrared Processing and Analysis Center, California Institute of Technology, Pasadena, CA 91125, USA}
\altaffiltext{20}{Harvard-Smithsonian Center for Astrophysics,
        60 Garden Street, Cambridge, MA 02138}
\altaffiltext{21}{Institute for Astronomy, University of Hawaii, 2680
  Woodlawn Drive, Honolulu HI 96822}
 \email{ariel@fysik.su.se}

\begin{abstract}
  We report on the discovery of SN~2014J in the
  nearby galaxy M~82. Given its proximity, it offers the best
  opportunity to date to study a thermonuclear supernova over a wide
  range of the electromagnetic spectrum. Optical,
  near-IR and mid-IR observations on the rising lightcurve, orchestrated by the
  intermediate Palomar Transient Factory (iPTF), show that SN~2014J is
  a spectroscopically normal Type Ia supernova, albeit exhibiting high-velocity features in its
  spectrum and heavily reddened by
  dust in the host galaxy.  Our earliest detections start just
  hours after the fitted time of explosion.  We use
  high-resolution optical spectroscopy to analyze the dense
  intervening material and do not detect any evolution in the resolved
  absorption features during the lightcurve rise. Similarly to other
  highly reddened Type Ia supernovae, a low value of
  total-to-selective extinction, $R_V \lsim 2$, provides the best
  match to our observations.  We also study pre-explosion optical and
  near-IR images from HST with special
  emphasis on the sources nearest to the SN location.
\end{abstract}


\section{Introduction}
Type Ia supernovae (SNe~Ia) are among the most luminous transient events at optical wavelengths
and extremely valuable tools to measure cosmological distances, see 
\citet{2011ARNPS..61..251G} for a recent review. 
Yet, SNe~Ia close enough to allow for detailed scrutiny of their physical properties
are very rare, especially in a galaxy like M~82, the 
host of several recent core-collapse supernovae \citep{2013MNRAS.431.2050M,2013MNRAS.431.1107G}.
At an estimated distance to M~82 of 3.5 Mpc
\citep{2009ApJS..183...67D}, SN~2014J is the closest identified Type
Ia SN in several decades, possibly rivaled by SN~1972E in NGC~5253 \citep{1973A&A....28..295A}
and SN~1986G in NGC~5128 \citep{1987PASP...99..592P}.  Thus, SN~2014J is exceptionally
 well-suited for follow-up observations in a wide range of
wavelengths, from radio to gamma-rays. These have the potential 
to yield transformational new clues into the progenitor systems of SNe~Ia, as
well as the detailed properties of dust along the line of sight, key
astrophysical unknowns for the study of the accelerated expansion of
the universe. 

There is strong evidence that SNe~Ia arise from thermonuclear
explosions of carbon-oxygen white dwarfs (WD) in binary systems \citep{2011Natur.480..344N,2012ApJ...744L..17B}. However, the nature of the
second star remains unclear. For a long time, the preferred 
scenario was the single degenerate (SD) model
\citep{1973ApJ...186.1007W}, where a WD accrets mass from a
 hydrogen or helium rich donor star, thus becoming unstable while approaching the Chandrasekhar mass. The double-degenerate (DD) model involving the merger of
two WDs \citep{1981NInfo..49....3T,1984ApJS...54..335I,1984ApJ...277..355W} has
gained considerable observational support in recent years, see e.g.,
\citet{2012NewAR..56..122W}.

In this work, we search for potential signatures of a SD progenitor system, such as
variable Na~D lines, precursor nova eruptions, features in the early lightcurve, radio emission,
or a coincident source in pre-explosion in HST images.


\section{Discovery and classification}
\label{detection}
SN~2014J was discovered by \citet{Fossey} in $BVR$-band
images of M~82 obtained on 2014 January 21.81 UT 
at UCL's University of London Observatory.
We have performed image subtractions using pre-explosion
data from the Palomar P60 telescope as reference, calibrated with
nearby stars listed in the
APASS catalog\footnote{http://www.aavso.org/apass} 
yielding a discovery
magnitude of $R=10.99\pm0.03 $\,mag. The discovery image 
(S.~Fossey, private communication) and the P60 reference
image, as well as the difference between the two are shown in
Fig.~\ref{fig:hstnir}, along with the pre-explosion HST images (GO:11360, PI: R.~O'Conell;
GO:10776, PI: M.~Mountain). The relative
position of SN~2014J with respect to neighboring stars (middle panel) was established using multiple short exposures in
$K$-band with adaptive optics and the NIRC2 wide camera at Keck
\citep{ATEL5789}. In Section~\ref{progenitor} we present a detailed
analysis of the pre-explosion data.

\begin{figure}
\plotone{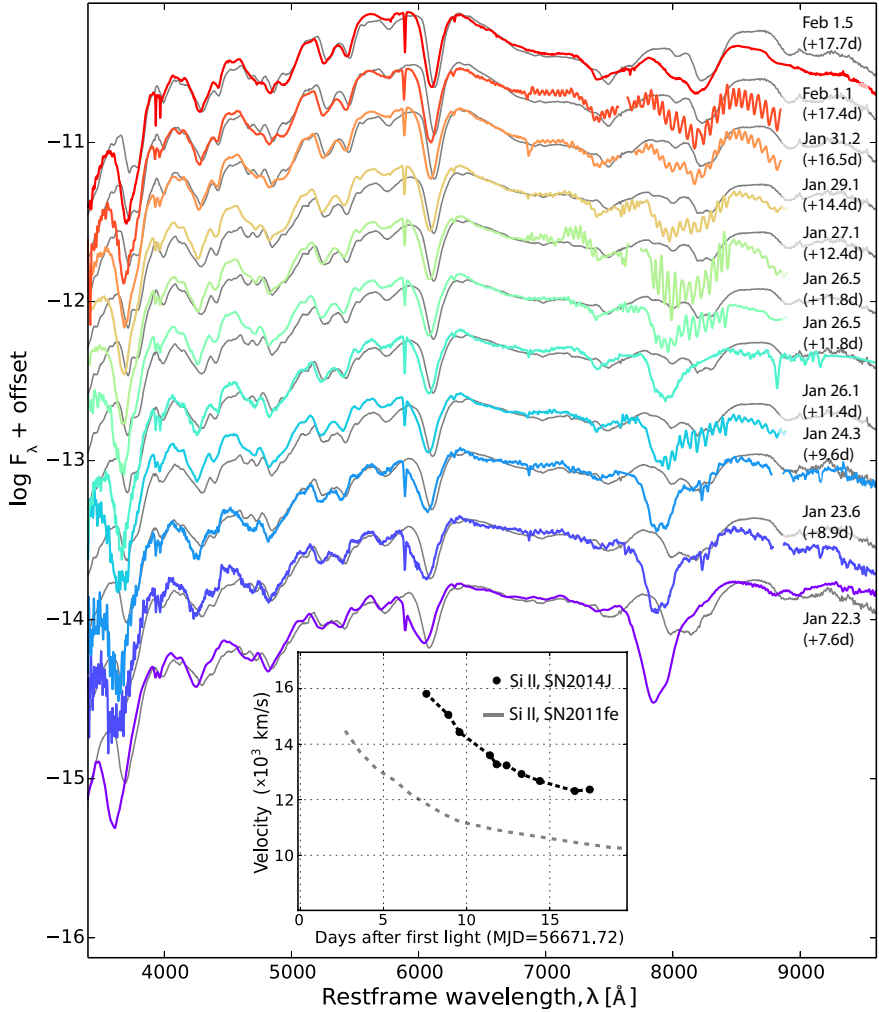}
  \caption{Pre-max spectroscopic follow-up of SN~2014J, starting 7.6 days
    from estimated supernova onset using the ARC 3.5m DIS, LCOGT FLOYDS, NOT
    ALFOSC and P200 DBSP spectrographs. Spectra of SN~2011fe \citep{2013A&A...554A..27P} at similar epochs (gray lines,
    reddened by $E(B-V)_{\rm host}=1.2$ mag with $R_V = 1.7$ and
    $E(B-V)_{\rm MW}=0.14$ mag with $R_V = 3.1$ are shown for comparison.  The
    inset panel shows that SN~2014J has higher
 SiII velocity than SN~2011fe and a steeper velocity gradient \citep{2005ApJ...623.1011B}.
\label{fig:lowres}}
\end{figure}

A classification spectrum was obtained by the iPTF team on January 22.30
with the Dual Imaging Spectrograph on the ARC 3.5m telescope
\citep{ATEL5786}, and in the near-IR using the MOSFIRE
instrument at Keck. The combined spectra are shown in
the bottom panel of Fig.~\ref{fig:hstnir}, while the photometry
collected so far is displayed in Fig.~\ref{fig:lc}.
 The object shows
characteristic spectral features associated with 
SNe~Ia, e.g., similar to SN~2011fe
\citep{2013A&A...554A..27P,2013ApJ...766...72H}.  However, the steep
attenuation of the spectrum at short wavelengths is indicative of
unusually large extinction by dust in the line of sight.  A good match to the 
overall SED 
is found invoking a pronounced color excess, $E(B-V)_{\rm host}
\approx 1.2$ mag, in addition to Galactic reddening, $E(B-V)_{\rm
  MW}=0.14$ mag \citep{2011ApJ...737..103S}, as shown in
Fig.~\ref{fig:lowres}. For the comparison, the spectra of SN~2011fe 
were artificially 
reddened assuming a Milky-Way type extinction law
\citep{1989ApJ...345..245C}, where both the color excess
and $R_V$ were allowed to vary freely.
The spectrum favors a low value of the total-to-selective extinction,
$R_V \lsim 2$, as also suggested by spectropolarimetry observations by
\citet{ATEL5830}.  Low values of R$_V$ are not unusual in SNe~Ia,
especially in the cases of high extinction, see
e.g.~\citet{2008A&A...487...19N}.  

\section{The iPTF-led multi-wavelength monitoring of SN~2014J in M~82}

As a part of its continuous survey of the sky in the search for
transients, iPTF has monitored M~82 since October
2009, with nearly daily cadence over the several months each year when
M~82 is visible from Palomar.  The most recent campaign started on
2013, November 28.  For the periods around full moon (e.g., around the
time SN~2014J exploded), not well suited for transient searches,
$H\alpha$ narrow-band imaging was conducted.
 
\begin{figure}
\epsscale{.70}
\plotone{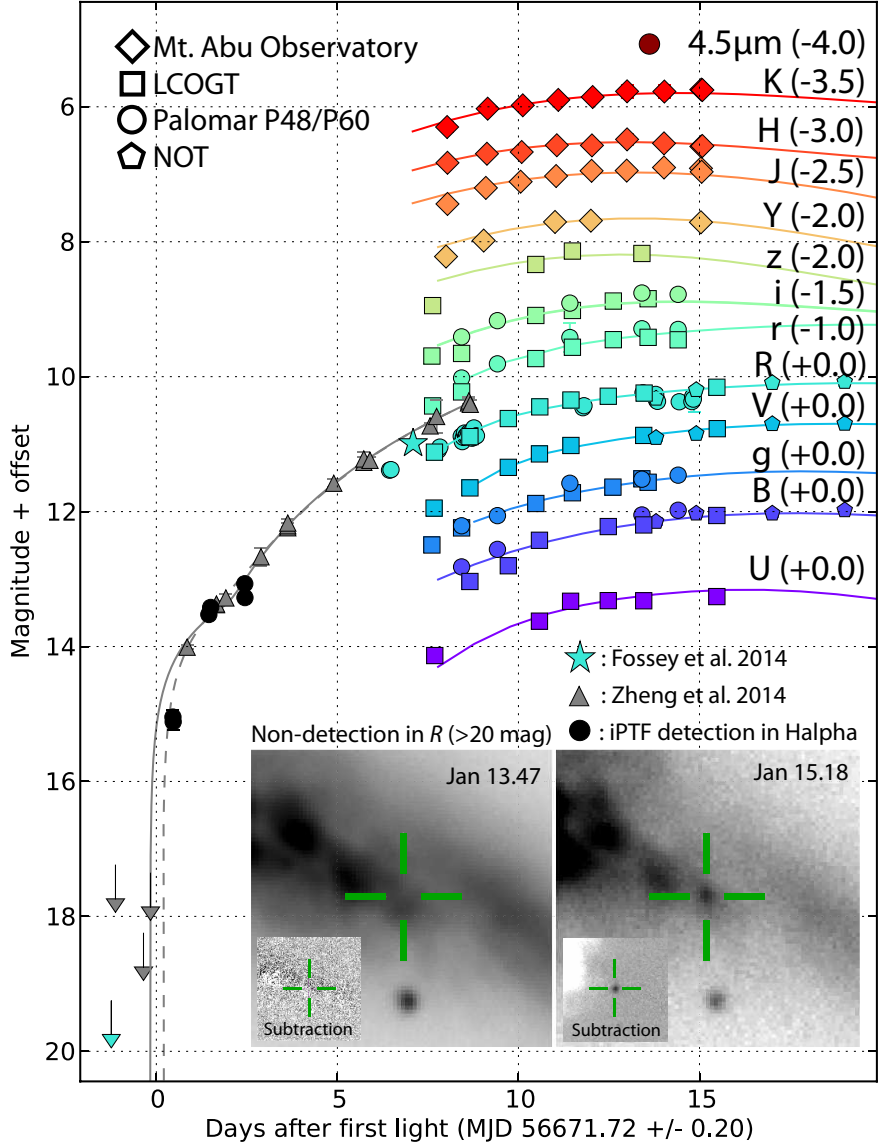}
\caption{Lightcurves showing the rise of SN~2014J, along with SNooPy fits described in the text. 
  The first P48 $H_{\alpha}^{656}$ and
  $H_{\alpha}^{663}$ narrow bands detections are shown 
  (black circles and inset image), S-corrected \citep{2002AJ....124.2100S} to $R$-band.
  Due to lack of accurate absolute calibration for the $H_\alpha$ filters, a common offset was applied to connect with the data-points
  in the pre-discovery light curve presented in
  \citet{2014arXiv1401.7968Z}. We also show their two fits of $t_0$,
  suggesting our first detection could be within $\sim$5 hours from the onset
  of the supernova.
\label{fig:lc}}
\end{figure}

The current best fit of the time of explosion, $t_0$,  was reported by KAIT
\citep{2014arXiv1401.7968Z} to be January 14.72 UT ($\pm$ 0.2 days).  
Upon later scrutiny of the pre-discovery P48 data, the supernova was found in several
observations from the iPTF $H\alpha$ narrow-band survey, starting just
hours after the fitted $t_0$.
We find a relative flux increase from January 15.18 to January 16.18 of 1.6 mag,  consistent with the ``method 2 fit'' in \citet{2014arXiv1401.7968Z}.  The SN
is also prominent on $R$-band photometry from the P48 prior to January 21
shown in Fig.~\ref{fig:lc}, but remained undetected 
by our automated software due to pixel saturation. 

Through an iPTF-led effort, involving also the Las Cumbres Observatory
Global Telescope (LCOGT) network \citep{2013PASP..125.1031B}, the Nordic Optical Telescope and the
Mount Abu Observatory \citep{ATEL5793}, we have secured optical,
near-IR and mid-IR lightcurves carefully monitoring the rise of the
supernova, as shown in Fig.~\ref{fig:lc}.  The $4.5\mu$m observations
were taken under the {\em Spitzer InfraRed Intensive Transients
  Survey} (SPIRITS; PI: Kasliwal).

The spectra shown in Fig.~\ref{fig:lowres} are consistent with
those from a normal Type Ia explosion, similar to e.g., SN
2011fe, but reddened following a CCM-law \citep{1989ApJ...345..245C}
with $E(B-V) \sim 1.2$ mag and $R_V=1.3-2$, in addition to
Galactic reddening.  
Fig.~\ref{fig:lc} also shows lightcurve fits using the SNooPy fitter
\citep{2011AJ....141...19B} of the photometric 
data prior to maximum brightness. 
Best fits are found for $E(B-V)_{\rm host}=1.22 \pm 0.05$
mag and $R_V= 1.4\pm0.15$. We expect the accuracy of the fitted parameters to 
improve, as the lightcurve shape estimate will profit from the
decreasing part of the SN lightcurve. However, the available data
clearly puts $R_V$ well below the Galactic average value,
$R_V=3.1$.

We also obtained two 1800 s high-resolution (R = 40000) spectra with
SOPHIE at Observatorie Haute-Provence on January 26.0 and January
28.0. Further, two 1800 s spectra (R = 67000) were obtained with the
FIbre-fed Echelle Spectrograph (FIES) on January 27.3 and another on February
1.0 with the Nordic Optical Telescope.

\begin{figure}
\epsscale{1.0}
\plotone{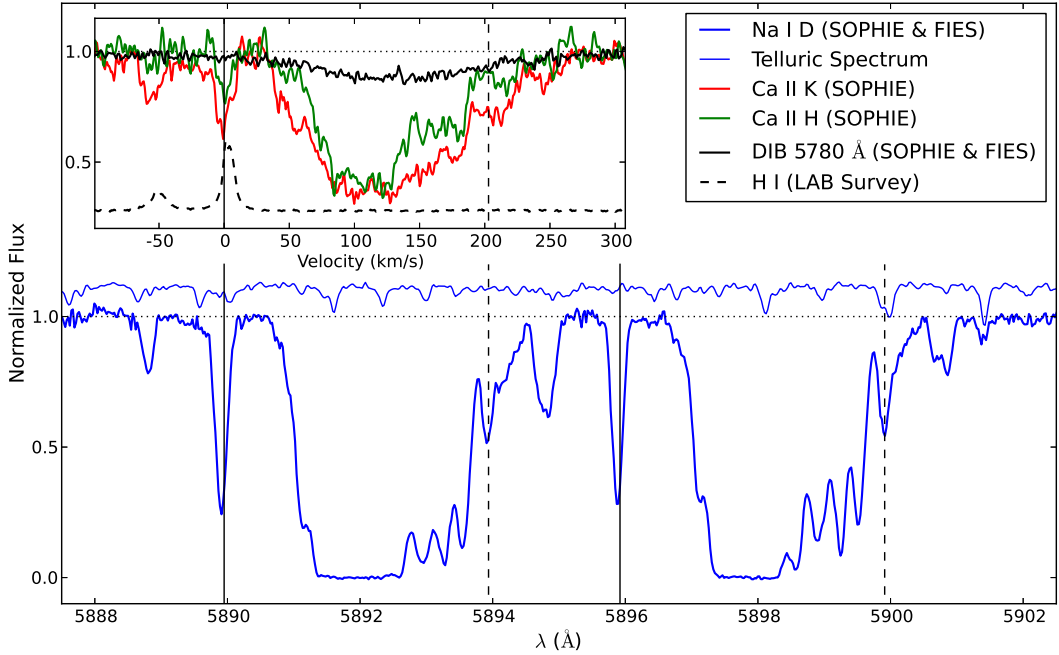}
\caption{The normalized Na~I~D doublet is plotted with solid vertical lines indicating the Galactic restframe wavelengths and the dashed vertical lines
  correspond to the mean velocity of M~82 (203 km\,s$^{-1}$). 
The inset panel shows the velocity distribution of Ca~II~H \& K (green and red lines) and the $\lambda$5780 DIB (black). 
Additionally, the H~I $\lambda$21 cm emission spectrum of the line of sight of M~82 from the LAB survey \citep{2005A&A...440..775K}
is plotted (dashed gray line).
The features of the Na~I~D and Ca~II~H \& K, at approximately -50 and 0  km\,s$^{-1}$ with respect to 
  the Galactic rest frame can be attributed to absorption in the Milky Way.
  The main Na~I~D absorption features, saturated between 74--135 km\,s$^{-1}$, originate in M~82.\label{fig:highres}}
\end{figure}

All spectra reveal deep multiple component Na~I~D absorption and
diffuse interstellar bands (DIBs), including $\lambda \lambda$
5780,5797,6284, and 6614, also reported by \citet{ATEL5797} and
\citet{ATEL5816}. The SOPHIE spectra further contain well resolved
Ca~II~H \& K with features matching those of the Na~I~D lines, shown in
the co-added spectrum in Fig~\ref{fig:highres}. We have not detected
any significant time evolution for any of the resolved components of
the Na~I~D doublet over the four epochs (at the $\sim$10\% level for 
$3\sigma$), thus motivating the
combination of the spectra.

Following the procedure outlined by \citet{2013ApJ...779...38P},  we measure the 
equivalent width (EW) of the $\lambda$5780 DIB to derive an independent
estimate of host galaxy extinction for SN~2014J. We find 
EW(5780) = 0.48 $\pm 0.01$~\AA \, corresponding to $A_{V}^{\rm host}=2.5\pm 1.3$ mag. 


Given the low recession velocity of M~82, it is difficult to separate the contribution from the Milky-Way and the SN host galaxy absorption. 
However, the availability of H~I data from the LAB survey\footnote{http://www.astro.uni-bonn.de/en/download/data/lab-survey/}  in the direction of M~82 
(see inset panel in  Fig.~\ref{fig:highres}; 
\citet{2005A&A...440..775K})  clearly indicates which features are Galactic.
Hence, all the absorption features redshifted with respect to the Milky Way are due to intervening material in M~82. 


\section{Spectral modeling}
\label{spec}

In Fig.~\ref{fig:lowres} we present a time-series spectral comparison
between SN~2011fe and SN~2014J starting roughly 12 days before peak
brightness. The SNe are remarkably similar in their spectral
evolution. The main differences seen are that the overall velocities
are higher in SN~2014J (see the inset \ion{Si}{2} velocity plot) and
there is a strong signature of high-velocity \ion{Si}{2} and
\ion{Ca}{2} in this supernova. 

\begin{figure}
\epsscale{1.0}
\plotone{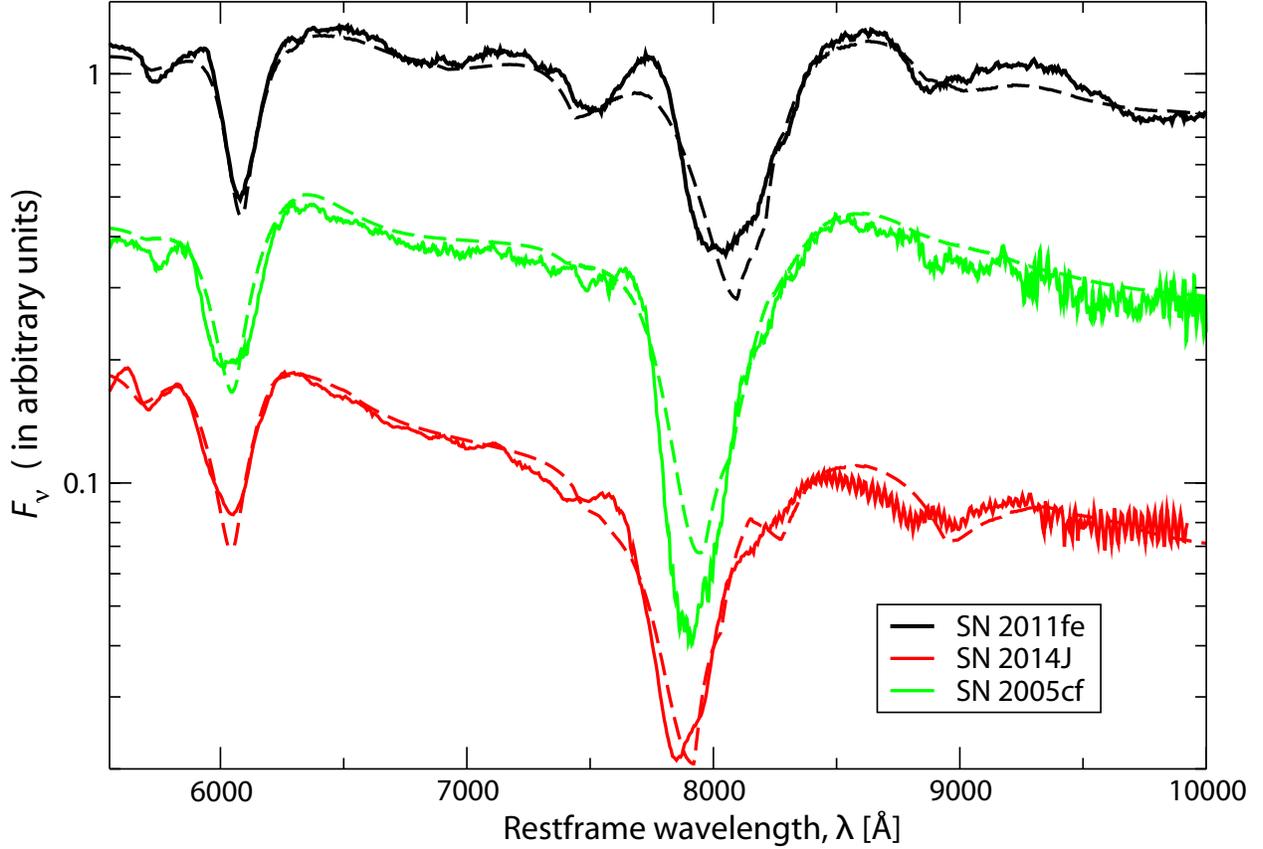}
\caption{Early spectra of SNe~2011fe, 2005cf and
  2014J, concentrating on the red-side of the optical spectra,
  along with SYNAPSS \citep{2011PASP..123..237T} fits to each (dashed
  lines). In these fits we have employed the ions: C~II, O~I, Mg~II,
  Si~II and Ca~II with the latter two having both photospheric and
  high-velocity components. These fits highlight the major difference
  between these supernovae. SN~2014J shows no sign of either C~II or
  O~I in this part of the spectrum and, unlike 2011fe, quite strong
  high-velocity components of Si~II, and Ca~II extending well beyond
  20,000 km\,s$^{-1}$. In addition, the Mg~II photospheric feature in SN~2014J
  is stronger as well at this phase. SN~2005cf shows similar
  high-velocity Si~II and Ca~II to SN~2014J, but differs from this
  supernova as photospheric C~II is clearly seen at this phase and our
  SYNAPPS fits also favor O~I at this phase. This suggests that the
  nuclear burning in the outer layers of SN~2014J was more complete
  than that of SN~2011fe and the lack of C~II compared to SN~2005cf
  may imply the same or perhaps a viewing angle effect due to an
  off-center detonation
  \citep{2011ApJ...732...30P}.\label{fig:synapps}}
\end{figure}

To further investigate these differences we carried out a set of
SYNAPPS \citep{2011PASP..123..237T} fits to these two supernovae as
well as to SN~2005cf, which was distinct in its pervasive high-velocity
features \citep{2009ApJ...697..380W}. We present the results in
Fig.~\ref{fig:synapps}.  In our fits to the red-side of the optical
spectra we have employed the ions: C~II, O~I, Mg~II, Si~II and Ca~II
with the latter two having both photospheric and high-velocity
components. We see that SN~2014J more closely resembles
SN~2005cf with respect to the high-velocity \ion{Si}{2} and
\ion{Ca}{2} features which extend easily over the range of
20,000-30,000 km\,s$^{-1}$. Unlike either SN~2005cf or SN~2011fe, C~II and O~I
are absent at this phase in SN~2014J. We searched for the presence of
C~I~$\lambda$~1.0693~$\micron$ line in our NIR spectrum, but the
signal-to-noise is too low to do a meaningful fit comparable to that
done for SN~2011fe \citep{2013ApJ...766...72H}. We do note that due to
the stronger than average Mg~II features seen in the optical for
SN~2014J, this analysis may be more challenging, even with higher
quality spectra, as the presence of
Mg~II~$\lambda$~1.0927~$\micron$ should be quite strong as well.

Several explanations for the origin of the high-velocity
features have been presented, from density enhancements via swept up
CSM \citep{2004ApJ...607..391G,2006ApJ...645..470T} to mixing
or more complete burning in the outer layers of the supernova 
\citep{2005MNRAS.357..200M,2005ApJ...623L..37M}
to ionization effects
in the outer layers \citep{2013MNRAS.429.2127B}. What is clear is that
the features do offer a unique diagnostic for understanding properties
of the progenitor system and/or the explosion mechanism and
correlations between the strength of these features and the underlying
colors and lightcurves of the SNe~Ia \citep{2014MNRAS.437..338C,2013MNRAS.436..222M}.

SN~2014J is among a class of SNe Ia where high-velocity features are present yet little to no evidence for
C~II exists even in very early spectra (see the broad line or high-velocity gradient examples in \citet{2011ApJ...732...30P}). 
Since extensive UV data from HST exists for both SNe 2005cf and 2011fe, it will be interesting to see which of these supernovae
SN~2014J most closely matches with respect to both the color and luminosity evolution.


\section{The quest for the progenitor system}
\label{progenitor}
\begin{figure}
\epsscale{0.83}
\plotone{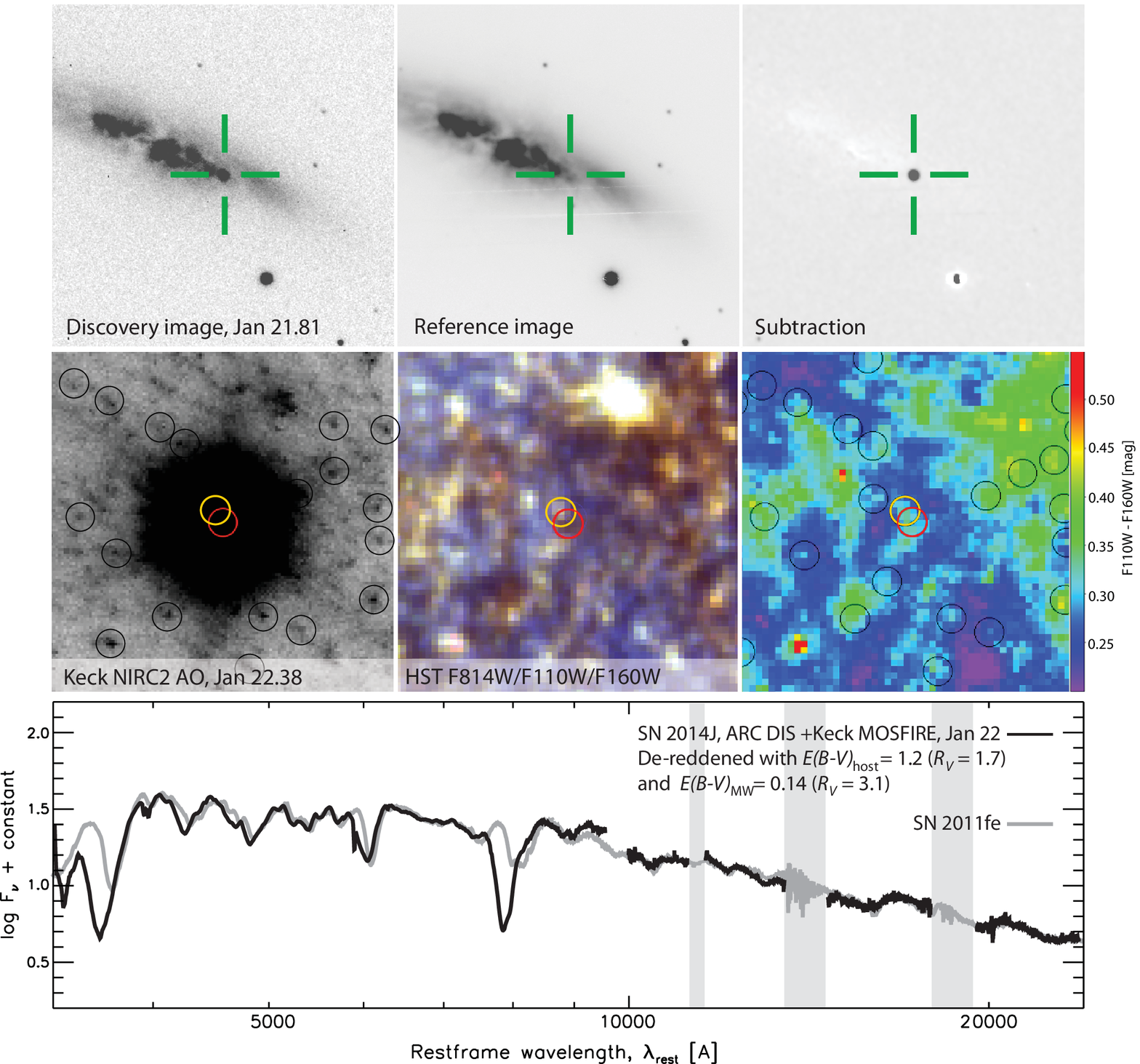}
\caption{Top panels: $5\arcmin \times 5\arcmin$ sections of the discovery \citep{Fossey}, reference and the subtraction images.  Middle panels (from left to right): $5\arcsec \times 5\arcsec$ section of the Keck
  NIRC2 adaptive optics image used to match the SN coordinates (red circle)
  to the surrounding sources (black circles). 
The nearest resolved object (yellow circle) in the pre-explosion HST composite images
  ($F815W/F110W/F160W$) is offset by $0.2\arcsec$ from the best estimate of the SN position.
The middle-right panel shows a color map 
  ($F110W-F160W$) indicating the large scale structures, probably due to patches of dust.
 The bottom panel shows the first optical and NIR spectrum of
  SN~2014J and a comparison to a combined spectrum from SN~2011fe by
  \citet{2013A&A...554A..27P} and \citet{2013ApJ...766...72H}, described in the text.
 \label{fig:hstnir}}
\end{figure}

M~82 has been extensively imaged by HST, thus it is possible to study
the environment of the SN prior to the explosion. Because of
the large attenuation due to dust in the line of sight, we concentrate
on the NIR bands. We
perform aperture photometry on the nearest sources to SN~2014J, shown
in Fig.~\ref{fig:hstnir}. The closest object (yellow circle in Fig.~\ref{fig:hstnir}) falls 0.2$\arcsec$ from
the current best estimate of the SN location
\citep[RA = $9^{h} 55^{m} 42^{s}.217(1)$, 
Dec = $69^{\circ} 40{\arcmin} 26{\arcsec}.56(4)$ in J2000 coordinates with respect
to the HST image,][]{ATEL5789}, corresponding to a $4\sigma$ spatial offset.
For this source, we measure AB magnitudes of $F110W = 21.4 \pm 0.4$,
$F128N = 21.9 \pm 0.4$, $F160W = 20.6 \pm 0.4$, $F164N = 21.2 \pm
0.4$.  The error is dominated by the uncertain background subtraction
as a result of source confusion.  The $F110-F160W$ color is typical of
other sources near this position. 
At a distance of 3.5 Mpc, the corresponding absolute magnitude of the
nearest resolved object is $J_{\rm AB} \sim H_{\rm AB} \sim -7$ mag ($A_H<A_J<0.4$ mag).  This source
could represent a stellar cluster, a grouping of unrelated objects 
or a region of
relatively low dust attenuation.

Next, we consider the possibility that the source is a donor star in
the SD scenario. 
The derived luminosity
would then suggest a very luminous red supergiant.  However, in the
case where a SN originates from a system with CSM created by a mass-loss from a donor star, the interaction
between the SN ejecta and the CSM is expected to give rise to radio
emission \citep{1982ApJ...259L..85C,1988Natur.332..514C}. The radio
null-detections on January 23  and 24 \citep{ATEL5800,ATEL5812} can therefore be used to derive an
upper limit on the mass-loss rate. Adopting a SN shock-wave velocity of
$3\times 10^{4}\,{\rm km\,s^{-1}}$ (about twice the Si II velocity, see
Fig.~\ref{fig:lowres}), and similar parameters as assumed by
\citet{2012ApJ...746...21H} for SN\,2011fe, the upper limit on the
mass-loss rate is $\dot{M} \leq 7 \times 10^{-9}(w/100\,{\rm km
  \,s}^{-1})\,M_{\odot}\,{\rm yr}^{-1}$, where $w$ is the mass-loss
wind velocity. 
The upper limit is comparable to the ones obtained for SN\,2011fe \citep{2012ApJ...746...21H,2012ApJ...750..164C}.  Given these tight limits and the
spatial displacement,
we conclude that the closest resolved source in the pre-explosion HST
images is unlikely to be a donor star.

Finally, we note that the $F110W-F160W$ color
map shown in the middle right panel of Fig~\ref{fig:hstnir} suggests that SN~2014J is at
the edge of a dust patch, about 4 pc in projected size. Light echoes may
thus be expected for this supernova.

We have searched for possible nova outbursts in the historic P48
$R$-band data covering a period of about 1500 days prior to the
detection of SN~2014J. By binning the data in bins of 15 days, we do
not find any excess larger than $4\sigma$
(calculated using the bootstrap technique, see 
\citet{1982jbor.book.....E}).  Our limiting magnitude is $R>19.5$\,mag
for a total time span of 1000\,days, and $R>20.25$\,mag for more than
765\,days in this 1500\,days time window.  Assuming $A_R=2$ mag,
compatible with the extinction we estimate based on the supernova colors, 
this corresponds to absolute
magnitudes $M_{{\rm R}} =-10.2$ and $-9.5$ respectively at the
distance of M~82. However, given the uncertainties on the properties of
recurrent novae, see e.g.~\citet{2014arXiv1401.2426T}, we refrain from
drawing firm conclusions against the possibility of recurrent novae
preceding SN~2014J based on these non-detections.

\section{Summary and conclusions}
The discovery of SN~2014J presents us with a unique opportunity to
explore the physics of Type Ia SNe and the line-of-sight effects due
to intervening matter. Further understanding in these areas is of
utmost importance for the use of SNe~Ia in cosmology.  The early data
from P48, starting as early as only hours from the explosion, and the
multi-wavelength follow-up by the iPTF team covers an important range
of the available windows in the electromagnetic spectrum.  Just as the
lightcurve reaches its maximum, we have learned that the SN has
suffered non-standard extinction. We have searched for, but not
detected, any time variation in our high-resolution spectra of the
Na~I~D doublet. Similarly, we do not detect any pre-explosion activity
in the $\sim$1500 days of P48 monitoring. In a study of pre-explosion
HST images in the near-IR, the nearest resolved source is found
$0.2{\arcsec}$ away from the SN location. The source brightness and
offset from the SN makes it unlikely as a donor star in a
single-degenerate scenario.

Further, we make a first study of the spectral features of SN~2014J
and find that it exhibits high-velocity features from intermediate
mass material but lacks C and O often seen in very early
spectra. Otherwise, it is a very similar to several well-studied
normal SNe Ia.







\acknowledgments 
We thank S.~Fossey for making the discovery $R$-band image of SN~2014J available to us. We are grateful to S.~Fossey and M.~Phillips for helpful comments on the manuscript. 
We acknowledge A. McKay, A. Bradley, N. Scoville, P.~L.~Capak,
C.~M.Carollo, S.~Lilly, H.~Sheth,
V.~Bhalerao, P.~Donati, S.~Geier, F.~Saturni, G.~Nowak
and A.~Finoguenov for cooperating with ToO
observations. AG and RA acknowledge support from the Swedish Research
Council and the Swedish Space Board and A.~G.-Y an ERC grant. MMK acknowledges generous
support from the Hubble Fellowship and Carnegie-Princeton Fellowship.
Based on observations made with the Nordic Optical Telescope, operated
by the Nordic Optical Telescope Scientific Association at the
Observatorio del Roque de los Muchachos, La Palma, Spain, of the
Instituto de Astrofisica de Canarias, Faulkes Telescope North image
observed by Gain Lee, the Mount Abu 1.2m Infrared telescope,
India and the
1.93m telescope of Haute-Provence Observatory, CNRS, France.
This research used resources of the National Energy Research
Scientific Computing Center, which is supported by the Office of
Science of the U.S. Department of Energy under Contract
No. DE-AC02-05CH11231.


\bibliographystyle{apj}

\clearpage

\end{document}